\documentclass[conference]{IEEEtran}
\IEEEoverridecommandlockouts

\usepackage{cite}
\usepackage{amsmath,amssymb,amsfonts}
\usepackage{algorithmic}
\usepackage{graphicx}
\usepackage{subcaption}
\usepackage{multicol, multirow}
\usepackage{textcomp}
\usepackage{xcolor}
\usepackage{url}
\usepackage{todonotes}
\usepackage{makecell}
\makeatletter
\newcommand*{\rom}[1]{\expandafter\@slowromancap\romannumeral #1@}
\makeatother
\def\BibTeX{{\rm B\kern-.05em{\sc i\kern-.025em b}\kern-.08em
    T\kern-.1667em\lower.7ex\hbox{E}\kern-.125emX}}
\begin{document}

\title{"Good" and "Bad" Failures in Industrial CI/CD -- Balancing Cost and Quality Assurance
\thanks{This research has been partially financed by Software Center, www.software-center.se}
}


\author{\IEEEauthorblockN{Simin Sun}
\IEEEauthorblockA{\textit{Chalmers University of Technology} \\ \textit{and University of Gothenburg} \\
Gothenburg, Sweden \\
simin.sun@gu.se}

\and

\IEEEauthorblockN{David Friberg}
\IEEEauthorblockA{\textit{Zenseact} \\
Gothenburg, Sweden \\
david.friberg@zenseact.com}

\and

\IEEEauthorblockN{Miroslaw Staron}
\IEEEauthorblockA{\textit{Chalmers University of Technology} \\ \textit{and University of Gothenburg} \\
Gothenburg, Sweden \\
miroslaw.staron@gu.se}
}

\maketitle

\begin{abstract}
Continuous Integration and Continuous Deployment (CI/CD) pipeline automates software development to speed up and enhance the efficiency of engineering software. These workflows consist of various jobs, such as code validation and testing, which developers must wait to complete before receiving feedback. The jobs can fail, which leads to unnecessary delays in build times, decreasing productivity for developers, and increasing costs for companies. 

To explore how companies adopt CI/CD workflows and balance cost with quality assurance during optimization, we studied 4 companies, reporting industry experiences with CI/CD practices.

Our findings reveal that organizations can confuse the distinction between CI and CD, whereas code merge and product release serve as more effective milestones for process optimization and risk control. While numerous tools and research efforts target the post-merge phase to enhance productivity, limited attention has been given to the pre-merge phase, where early failure prevention brings more impacts and less risks.
\end{abstract}


\section{Introduction}
In modern software development, Continuous Integration and Continuous Delivery (CI/CD) pipelines are essential for improving development efficiency and maintaining software quality. Since the introduction of these concepts by Fowler and Foemmel in 2006~\cite{fowler2006continuous}, a wide range of CI/CD tools—such as Jenkins, Travis CI, GitLab CI, and GitHub Actions—have emerged to automate the process~\cite{cruisecontrol, jenkins, travisci, gitlabci, githubactions}. A typical pipeline involves committing code, triggering builds via a CI server, receiving automated feedback, and updating the repository~\cite{hilton2017trade}.

CI/CD pipelines execute numerous jobs to enforce testing, security, and compliance standards. However, developers frequently face long build times and delayed feedback, which reduce productivity and increase operational costs. To address this, researchers have proposed a variety of optimization techniques, including job prioritization~\cite{abdalkareem2019commits, jin2022builds}, selective execution~\cite{jin2023hybridcisave}, and prediction-based methods for anticipating job outcomes~\cite{bisong2017built, al2022predicting, saidani2022improving}. Despite these advances, challenges remain in balancing efficiency with risk—early failures that are missed can lead to larger issues downstream~\cite{hilton2017trade}.


While the adoption of CI/CD research has evolved in the last decade, the ways in which pipelines are viewed and implemented in practice have also evolved. In particular, organizations are increasingly recognizing the need to optimize pipelines not only for performance but also for reliability and maintainability. In this paper, we explore how contemporary software teams manage their CI/CD workflows, with a focus on how they balance cost and quality considerations. By engaging directly with developers and CI/CD practitioners across different organizational contexts, we examine the coexistence, migration, and optimization of CI/CD tools in real-world environments.

Our empirical investigation is grounded in a year-long study involving eight participants from four software companies, ranging from large-scale enterprises to small startups. These collaborators provided in-depth insights into their development practices, tooling architectures, and strategic decisions around CI/CD adoption and evolution. Their contributions were instrumental not only in shaping the direction of our research but also in identifying key challenges and opportunities. This study aims to answer the following two research questions:

\begin{itemize}
    \item RQ1: What are the current CI/CD architectures used by the surveyed companies, and what factors influence their architectural choices?
    \item RQ2: How can CI/CD jobs be categorized to balance the trade-off between higher productivity and lower risk?
\end{itemize}

Our findings show that modern CI/CD pipelines can be divided into three distinct phases, separated by two critical milestones: code merge and product release. The dynamics of the CI/CD pipelines change significantly in these parts and reflect a shift in ownership, responsibility, and impact. While the product release has long been recognized as a transition in the pipeline, our results highlight the growing significance of the code merge phase as a pivotal transition. In the pre-merge phase, developers hold primary responsibility, and issues tend to affect individuals, often leading to personal frustration. In the post-merge phase, responsibility shifts toward the organization, and while more developers may be impacted by failures, individual accountability becomes diffuse.
We find that the pre-merge phase presents a unique opportunity for low-risk, high-impact improvements.


\section{Methodology}
We aimed to identify industrial practitioners with experience in CI/CD workflows, DevOps, or familiarity with these processes. To achieve this, we contacted our collaborating companies and asked them to shortlist potential candidates who met the criteria. We then contacted the shortlisted participants to assess their expertise and willingness to participate in the research. Eventually, we had discussions with eight experts from four different companies. The sample represents professional software engineering in the embedded software domain, covering a variety of organization contexts and sizes and practitioner roles.           

\subsection{Data Collection and Analysis}
We conducted this research by a series of discussions, either in person, via Microsoft Teams or by E-mail. Each time began with a brief introduction to our study, followed by obtaining participants’ consent to use the content of our conversation for research purposes.

The discussions followed a structured set of open-ended questions categorized into three main areas: (1) CI/CD pipeline, and (2) Optimization strategy. To better understand CI/CD practices, we inquired about tool usage, existing jobs, migration history, tool coexistence, and participants' motivations for these choices. We also explored the cost of their current CI/CD pipeline in terms of time and effort, as well as strategies for cost reduction, including build time optimization, skip/non-skip strategies, and test case prioritization. Participants were asked about the advantages and disadvantages of these strategies.

All participants were anonymized, and no personal data were collected or stored. Some participants provided real-world design representations under confidentiality agreements. Our analysis followed an inductive approach, with an initial classification based on CI/CD tools. To validate our findings, we presented the categorized data and analysis to an experienced software developer and a professor of software engineering for independent review. To analyze the qualitative data, we followed established guidelines~\cite{runeson2009guidelines}.

\section{Results}
\subsection{RQ1}
The first finding about the architectures of CI/CD in all participators' companies is that tool coexistence and migration are prevalent across all companies, often functioning as both a cause and a consequence of workflow changes. This observation aligns with prior research on tool migration~\cite{rostami2023usage} and the adoption of GitHub Actions~\cite{decan2022use}. GitHub Actions has gained popularity due to increasing industry demands for automation and its scalability in large-scale development. However, Jenkins remains widely used across all participated companies, primarily due to its flexibility, extensive plugin ecosystem, and ease of integration into existing workflows. We also noticed that all recent research done in the participated companies did their trail test using the Jenkins instead of the real CI/CD tool. 

\begin{figure}[!htb]
  \centering
  \includegraphics[width=\linewidth]{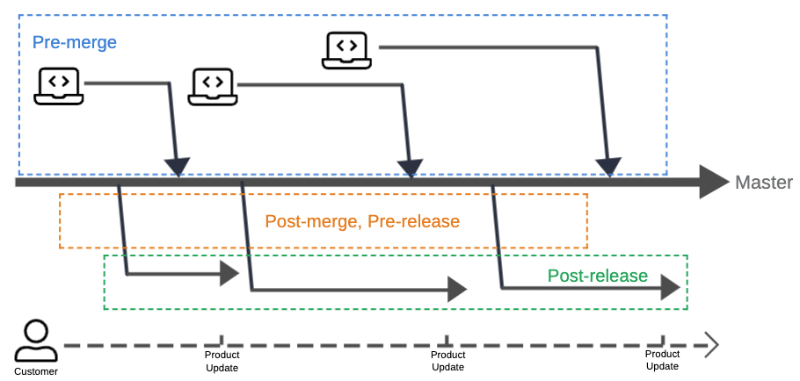}
  \caption{Phases in the CI/CD Pipeline}
   \label{fig:phase}
\end{figure}

The second finding comes from the discussion about different phases of CI/CD pipelines, all participants noted the difficulty of defining a clear milestone that marks the transition from CI to CD, particularly in companies providing software for larger systems. Additionally, they emphasized that CI activities may continue even after software release, as integration with other systems may be necessary. Unlike traditional CI/CD pipelines, where CI and CD are sequential, these processes often occur in parallel. However, we found the code merge and product release are clear milestones during the software development process, and we demonstrate it in Figure~\ref{fig:phase}. 

\begin{figure*}[h]
  \centering
  \includegraphics[width=\linewidth]{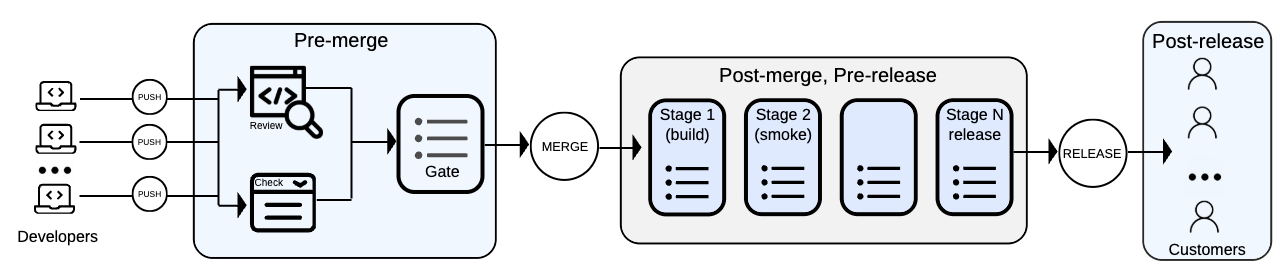} 
  \caption{Standard Pipeline Structure Observed Across Multiple Companies}
  \label{fig:arch}
\end{figure*}

Lastly, based on their descriptions, we aim to develop a general framework applicable to all participated companies, regardless of their domain or size. The common architectures we identified, which are widely used across these companies, are summarized in Figure \ref{fig:arch}.


\subsection{RQ2}
While CI/CD pipelines include a variety of jobs that differ across companies, some are widely adopted. According to our participants, these jobs can generally be categorized as follows:
\begin{itemize}
    \item Code Quality: Includes tasks such as static code analysis and functional test cases.
    \item Security: Includes security vulnerability checks.
    \item Dependency and Compliance: Involves verifying dependency versions and compliance verification across different software versions and system architectures
    \item License Management: Ensures the proper licensing of all packages used.
    \item Infrastructure and Configuration: Focuses on validating the configuration and related infrastructure.
    \item Deployment and Release: Includes mock installations and software deployment simulations.
\end{itemize}

The ideal or safest scenario is to run all jobs at the pre-merge phase to ensure software security and reliability. However, this approach can lead to inefficiencies and increased costs, as developers must wait longer for the pipeline to complete, ultimately lowering productivity and raising operational expenses. 

To better understand the trade-offs, we analyzed CI/CD job execution data from 2024 at one company. We found a notable discrepancy in failure rates—the failure rate of pre-merge checks compared to post-merge is 5:3. Moreover, this does not take into account that many post-merge failures are sticky, meaning they are counted repeatedly until a fix is applied and has propagated downstream. As a result, the ratio may understate the true discrepancy, since it conservatively overestimates the number of unique post-merge failures. We also found that the overall ratio of checks run in pre-merge versus post-merge is approximately 15:1. This highlights not only the imbalance in the total number of check runs between the pre-merge and post-merge phases, but also that pre-merge runs—already far more numerous—are more prone to failure. 
A key contributor to this higher failure rate is the significantly greater number of jobs per check in the pre-merge phase, compared to the number of jobs that are unique to post-merge—that is, jobs which have not already succeeded in pre-merge.

When considering the higher job count per check in pre-merge, the scale of check volume—with millions of runs annually—and the constrained resources within organizations, these findings suggest that optimizing the pre-merge phase—whether by reducing failures or improving efficiency—could drive more immediate and meaningful improvements to the developer experience than equivalent efforts focused on post-merge.


Another key observation is that, in addition to the commonly reported causes of long build times and job failures—such as test failures, build mis-configurations, and dependency issues, as noted in prior research \cite{ghaleb2019empirical, rausch2017empirical}—two of the participants highlighted flaky tests and intermittent failures as major contributors to CI/CD job failures in their companies. Notably, recent research \cite{aidasso2025diagnosis} has begun to focus on these types of failures, underscoring their growing significance in the field. Additionally, for one company, software configuration management (SWM) and product configuration management (PCM) were identified as key failure points, particularly in infrastructure- and configuration-related jobs. For their release process, ensuring compatibility between all system components, including both hardware and software, was crucial.

While long build times pose challenges for both developers and organizations, there is hesitation in adopting early-phase build optimization techniques, especially those involving skipped builds or test cases. Participants preferred exposing failures earlier rather than later, aiming to detect as many issues as possible in the early phases. By contrast, failures in later phases -- particularly during integration into a more extensive architecture or product release -- should be minimal or nonexistent. As a result, unless skipping techniques provide trustworthy and explainable justifications, they are used with caution. 

Currently, job prioritization is done using cache-based and code comparison techniques. Our analysis showed that teams rely on these tools to analyze diffs and dependencies in code to validate new commits efficiently and schedule the execution of jobs. However, they primarily operate in the post-merge and pre-release phases. 

While post-merge optimization remains a widely studied area in CI/CD research, there is growing interest in pre-merge optimizations. Since code at this phase has not yet been merged into the main branch, any issues identified can be addressed with minimal cost and no risk to the overall system architecture. Three of the four companies expressed interest in solutions for predicting build outcomes at runtime or during the pre-merge phase, as these approaches could significantly reduce resource consumption and prevent failures. Participants from two different companies mentioned that they had done research on exploring this direction using generative artificial intelligence models to enhance prediction accuracy and streamline CI/CD workflows.


\section{Discussion}
\label{sec:discussion}
\noindent
The findings of this study highlight two critical milestones: code merge and product release, in the software development lifecycle that are widely recognized across companies and developers. The implications of our research include the following:

\textbf{For company:} A key insight from this study is the distinction between "good" failures, which occur early in the development process (e.g., pre-merge), and "bad" failures, which arise at later phases (e.g., post-merge). The cost of addressing issues at these phases varies significantly. While companies aim to enhance developer productivity by reducing build frequency and optimizing build time in the early phases, they must also consider the risks associated with such strategies. A balance must be struck to prevent critical failures in later phases, which can be far more costly. 

A widely adopted approach is history-based analysis~\cite{lima2020test}, which leverages failure and test execution history to detect faults in submitted code. This remains an active research area. In industry, tools like Bazel have gained popularity for their ability to track history and perform comparisons. Another strategy is scheduling automated build jobs during off-peak hours, such as evenings or weekends, to minimize developer wait times and reduce computational load. These methods are generally safe, with skipping and auto-run mechanisms remaining under manual control.

While some companies have observed that postponing rarely failing jobs to later phases can improve overall efficiency, they approach this with caution. Post-merge failures could result in pre-merge effect such as merge stop, which impacts the entire team. In such cases, resolving issues after the merge can be significantly more time-consuming and complex, potentially outweighing the benefits of deferred execution.

\textbf{For individual developers:} Our findings suggest that while pipeline bottlenecks occur in both the pre-merge and post-merge stages, individual developers often feel the impact more acutely during the pre-merge phase. This is the phase they interact with most frequently and where they encounter productivity barriers such as job failures, extended wait times, and time-consuming debugging. These challenges are align with prior research~\cite{hilton2017trade, saroar2023developers}. The productivity bottleneck is very much present in pre-merge, particularly as a) many jobs fails (and not always "good"), and b) the majority of load on the CI machinery comes from pre-merge.

Current research primarily focuses on test case prioritization\cite{lima2020test, jin2023hybridcisave}, build-skipping techniques\cite{abdalkareem2019commits, saidani2021detecting, jin2022builds}, and build prediction~\cite{al2022predicting, saidani2022improving}. However, these approaches operate at a higher level, optimizing overall efficiency rather than directly addressing developers' pain points. Developers still experience delays while waiting for builds to complete and must debug issues themselves.

While existing tools and optimization strategies primarily target post-merge and pre-release phases, there is a lack of support for the pre-merge phase—an area where improvements could further enhance developer efficiency and software quality. In practice, the dominant approach in most companies remains manual error resolution, with developers fixing CI/CD issues independently or seeking help from colleagues via group chats or in-person discussions.

\textbf{For researchers:} 
Based on Shahin’s systematic review~\cite{shahin2017continuous}, most current research aimed at facilitating the CI/CD process focuses on six key areas, namely shorten the build time, visibility and awareness of the results, automated testing, detecting the faults, security and scalability, and dependability and reliability in CD. However, all six areas primarily address the post-merge and pre-release phases, with limited attention given to predicting failures or resolving issues before developers commit their code. Notably, none of these areas directly enhance developer efficiency, as companies tend to prioritize quality assurance over efficiency.

However, there is an underexplored opportunity that could benefit both companies and developers: the pre-merge phase. By predicting build outcomes and assisting developers in debugging before code is merged or even committed, we could strike a balance between quality and efficiency, addressing a long-standing dilemma in CI/CD.

Furthermore, despite the diverse architectures of CI/CD workflows in practice, there is a pressing need for a universal framework that can address these challenges across different development environments. Future research should explore predictive models and proactive solutions to enhance early-phase error detection and resolution.


\section{Validity evaluation}
\label{sec:validity}
\noindent
We discussed the internal, external, construct and external validity thereat that could affect our results used the framework by Wohlin~\cite{wohlin2012experimentation}.
A threat to \textbf{internal validity} arises from participant selection, as their experiences and perspectives shape the conclusions of this study, making them indicative rather than definitive. To mitigate this, we selected participants with substantial experience in CI/CD and DevOps. Additionally, we asked participants to provide concrete examples when making strong claims to ensure the reliability of their insights.
A threat to \textbf{external validity} stems from the limited scope of our study and the specific settings of our discussions. Categorizing and evaluating CI/CD practices across four companies may affect the generalizability of our findings. To address these concerns, we selected companies of varying sizes and domains to enhance the diversity and applicability of our findings. 
A threat to \textbf{construct validity} arises if participants do not fully understand the questions or if the questions are not entirely applicable to their company. If a participant is not entirely familiar with certain aspects, we invite an additional participant from the same company to ensure all questions can be properly understood and answered.
A threat to \textbf{conclusion validity} arises from potential researcher bias in data analysis. Since this is a qualitative study, all conclusions are drawn from conversations with participants and analyzed based on meeting notes, which may introduce bias or lead to the omission of key insights. To mitigate this risk, we implemented an expert review process involving professionals from both industry and academia to ensure a more objective and comprehensive analysis.

\section{Conclusion}
\label{sec:conclusion}
\noindent

This empirical study highlights the critical role of CI/CD in modern software development and release processes. Practitioners take a broad view of the boundary between CI and CD, yet they consistently recognize two key milestones: code merge and product release. These milestones are frequently used to separate different phases of development, suggesting their potential application in CI/CD workflows to balance efficiency and risk management.

Our findings provide clear evidence that practitioners prefer "good" failures—those occurring early in the development process (e.g., pre-release)—over "bad" failures that emerge later (e.g., post-release). When considering optimization strategies, companies must weigh the trade-offs between efficiency gains and potential risks. While techniques such as build skipping and test case prioritization can enhance productivity, many companies hesitate to adopt them due to concerns about undetected failures. Instead, they favor strategies that provide clear explanations and traceability.

Additionally, our study identifies a significant gap in existing CI/CD tooling: pre-merge support. Current research primarily focuses on reducing build time after code submission and job execution, yet few tools help developers detect issues before merging code. Companies are actively seeking solutions that minimize waiting time while mitigating risks in early development phases or at runtime. Addressing this gap presents an opportunity for future research and tool development in CI/CD optimization.

In future studies, we aim to investigate developers' perspectives on pre-merge support in CI/CD, focusing on the specific tasks or areas where they seek assistance. Additionally, we plan to explore the potential of large language models in supporting these tasks and examine developers' opinions regarding their adoption in real-world practice.

\section{Acknowledgment}

We sincerely thank all industry participants for their valuable insights and contributions, which have been instrumental in shaping this work.

Software Center has also partially funded this work, a collaboration between the University of Gothenburg, Chalmers, and 18 universities and companies -- \url{www.software-center.se}.

\bibliographystyle{IEEEtran}
\bibliography{IEEEabrv,reference}

\end{document}